\documentclass[pdflatex,sn-mathphys-num]{sn-jnl}

\usepackage{graphicx}%
\usepackage{multirow}%
\usepackage{amsmath,amssymb,amsfonts}%
\usepackage{amsthm}%
\usepackage{mathrsfs}%
\usepackage[title]{appendix}%
\usepackage{xcolor}%
\usepackage{textcomp}%
\usepackage{manyfoot}%
\usepackage{booktabs}%
\usepackage{algorithm}%
\usepackage{algorithmicx}%
\usepackage{algpseudocode}%
\usepackage{listings}%

\theoremstyle{thmstyleone}%
\theoremstyle{thmstyletwo}%

\theoremstyle{thmstylethree}%

\begin{document}

\title[Article Title]{Efficient Nudged Elastic Band Method using Neural
Network Bayesian Algorithm Execution}


\author*[1]{\fnm{Pranav} \sur{Kakhandiki}}\email{pkakhand@stanford.edu}

\author[2]{\fnm{Sathya} \sur{Chitturi}}\email{chitturi@alumni.stanford.edu}

\author[2]{\fnm{Daniel} \sur{Ratner}}\email{dratner@slac.stanford.edu}

\author*[2]{\fnm{Sean} \sur{Gasiorowski}}\email{sgaz@slac.stanford.edu}

\affil*[1]{\orgdiv{Department of Applied Physics}, \orgname{Stanford University}, \orgaddress{\city{Stanford}, \postcode{94305}, \state{CA}, \country{USA}}}

\affil[2]{\orgname{SLAC National Accelerator Laboratory}, \orgaddress{\city{Menlo Park}, \postcode{94025}, \state{CA}, \country{USA}}}


\abstract{The discovery of a minimum energy pathway (MEP) between metastable states is crucial for scientific tasks including catalyst and biomolecular design. However, the standard nudged elastic band (NEB) algorithm requires hundreds to tens of thousands of compute-intensive simulations, making applications to complex systems prohibitively expensive. We introduce Neural Network Bayesian Algorithm Execution (NN-BAX), a framework that jointly learns the energy landscape and the MEP. NN-BAX sequentially fine-tunes a foundation model by actively selecting samples targeted at improving the MEP. Tested on Lennard-Jones and Embedded Atom Method systems, our approach achieves a one to two order of magnitude reduction in energy and force evaluations with negligible loss in MEP accuracy and demonstrates scalability to $>$100-dimensional systems. This work is therefore a promising step towards removing the computational barrier for MEP discovery in scientifically relevant systems, suggesting that weeks-long calculations may be achieved in hours or days with minimal loss in accuracy.}




\maketitle

\section*{Introduction}\label{sec1}

Understanding how materials and chemical systems change from one metastable state to another is crucial in predicting and understanding their behavior. Many such processes in physics, chemistry, materials science, and biology depend on the transition-state pathways (the sequence of atomic configurations connecting two metastable states) and the energy barriers (the energetic cost required to move along that path) that connect different states~\cite{piskulich2019activation}. These barriers dictate reaction rates, diffusion properties, and activation energies, which in turn affect macroscopic properties such as catalytic efficiency~\cite{schaaf2023accurate}. An example of a system governed by such energy barriers is conformational pathways in proteins~\cite{mathews2006nudged}, where structural changes are fundamental to biological function. Understanding these transitions lays the groundwork for controlling protein activity and developing synthetic biomolecules with tailored functions. Minimum energy pathways (MEPs) describe the most probable paths that atomic systems follow when transitioning between metastable states; critically, the lowest-energy pathway reveals the transition mechanism and corresponding energy barriers~\cite{eyring1935activated}. Nudged elastic band (NEB) is a popular method used for finding MEPs in atomic systems~\cite{jonsson1998nudged, henkelman2000improved, henkelman2000climbing}. In its basic form~\cite{jonsson1998nudged}, NEB begins with a set of ``moving images'' in atomic space between the initial and final state being studied, often using a linear interpolation as an initial guess. Force and energy calculations from physical simulations of the system are then used to nudge the band of images until a convergence criterion is met---commonly the maximum perpendicular force on any atom in any image, $f_{max}$, falling below a threshold. Each intermediate moving image is connected by an elastic spring force to ensure the band does not collapse to a single point. Since its development, various versions of NEB have been created to scale to more difficult systems or to reduce computational cost~\cite{trygubenko2004doubly, weinan2002string, maragakis2002adaptive, sheppard2012generalized, kolsbjerg2016automated}. However, a major drawback of these ``classical'' NEB methods is that they require many evaluations of potential energy functions, namely one per image for each NEB iteration. Such simulations may involve computationally expensive density functional theory (DFT) or molecular dynamics (MD) calculations. For example, a recent NEB study of catalysts \cite{wander2025cattsunami} with DFT took 52 GPU years for 19,000 NEB simulations. This time-consuming nature of NEB limits the possibilities of studying complex realistic systems, such as proteins and catalysts.

\begin{figure}[t]
\centering
\includegraphics[width=1.0\textwidth]{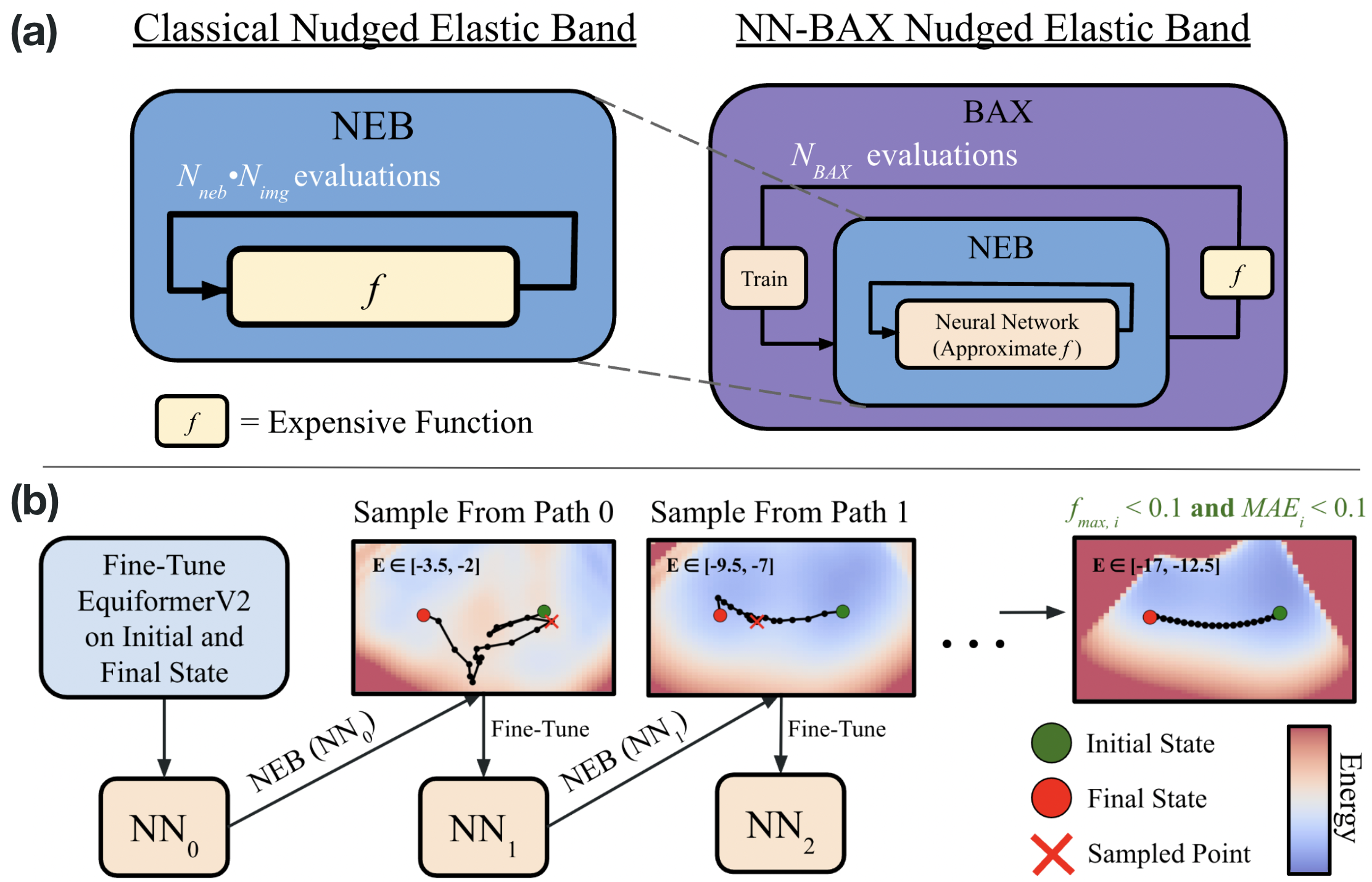}
\caption[BAX Pipeline]{NN-BAX Pipeline. a) Classical NEB versus NN-BAX NEB, where the expensive simulator \textit{f} is no longer called in the NEB Loop. The number of function evaluations in classical NEB scales with the number of images times the number of NEB iterations, $N_{neb} \cdot N_{img}$, whereas NN-BAX scales with the number of BAX iterations, $N_{BAX}$, need to learn the function. b) NN-BAX Loop. The EquiformerV2 model is initialized and trained on the initial and final state. In each loop, NEB runs using the trained model as an approximation for \textit{f}, a sample is acquired from the resulting path, and the sampled point is added to the training set for the subsequent iteration. Colormap energy ranges (eV) are provided.
}
\label{fig:1}
\end{figure}

Numerous efforts have been made to reduce the computational cost of NEB with machine learning~\cite{wander2025cattsunami, garrido2019low, koistinen2019nudged, fu2025learning, luo2025automatic, deng2025systematic, li2025transition, perego2024data, schaaf2023accurate, meyer2019machine, koistinen2017nudged, schreiner2022neuralneb, teng2024physical}. In one paradigm, neural network surrogate models are trained on large, diverse datasets. These models then serve as direct calculators for downstream applications, such as NEB calculations on complex systems. An alternative approach involves sequentially updating a simulation surrogate for a system of interest using techniques from active learning or Bayesian experimental design. Bayesian experimental design techniques can be used to infer properties of black-box functions with limited function queries. Common approaches fall into the Bayesian optimization (BO) family~\cite{jones1998efficient}, where the aim is to maximize or minimize the unknown function.  A useful tool in Bayesian experimental design is Bayesian algorithm execution (BAX)~\cite{neiswanger2021bayesian}, which generalizes the optimization objective to any computable function property (called the BAX ``algorithm''). However, these two existing methodologies, namely generic foundation models and Bayesian optimization, have limitations. Foundation models are fast, but sometimes not accurate enough on new systems and thus may not converge to the correct transition pathway. Bayesian experimental design approaches provide constant updates, but lack sophisticated models needed for complex systems. Specifically, previous work has used Gaussian processes (GPs) as a surrogate model, limiting their expressivity and scalability in high-dimensional spaces~\cite{binois2022survey}. The computational cost of GPs also scales cubically with dataset size~\cite{williams2006gaussian}, thus making GPs unfit for larger datasets. The approach in ~\cite{garrido2019low}, which fits into the BAX framework and uses a GP, tests on a maximum of 7-atom systems. 

In our work, we fuse these approaches, taking advantage of the development of large neural network surrogates for materials systems and applying them within the BAX acquisition framework, allowing for applicability to higher dimensional systems. Specifically, with BAX, we choose function queries that target the output of the NEB optimization on the unknown function defined by the physics simulator. We refer to this neural network BAX methodology as NN-BAX. There exist many foundation models for materials systems~\cite{riebesell2025framework}, and we use EquiformerV2~\cite{liao2023equiformerv2} as our surrogate model, pretrained on the OMat24~\cite{barroso2024open} dataset. The pre-trained EquiformerV2 model does not perform well for zero-shot NEB calculations on the systems we test. We therefore treat it as a foundation model and tune it in a few-shot-learning manner using BAX. A visualization of the pipeline is shown in Figure~\ref{fig:1}.

\section*{Results}\label{sec2}
We evaluate NN-BAX across a hierarchy of atomic transition systems of increasing physical realism and complexity, as described by two well-known empirical potentials. First we study transitions in Lennard–Jones clusters with 7 and 38 atoms, which provide well-characterized, high-dimensional testbeds with known minimum energy pathways. We then extend our analysis to a many-body Embedded Atom Method (EAM) potential for copper surface diffusion, demonstrating applicability beyond pairwise interactions. Finally, we consider complex multi-step transitions in 38-atom Lennard-Jones clusters and introduce Foundation-BAX, which leverages information shared across related paths to further reduce computational cost. Together, these systems allow us to assess the accuracy, efficiency, and scalability of NN-BAX across increasingly challenging regimes. 

For each NN-BAX iteration, we initialize the model to a pretrained foundation model. For the cases presented below, we consider the 153 million parameter EquiformerV2 trained on the OMat24 dataset. EquiformerV2 combines the Transformer architecture~\cite{vaswani2017attention} with consideration of relevant symmetries for materials systems: atomic energies are modeled as invariant under global translations and rotations of the system, while atomic forces are modeled as equivariant, transforming consistently with rotations and translations of the atomic positions. EquiformerV2 has demonstrated strong performance in benchmarks~\cite{riebesell2023matbench}
and previous non active learning (zero-shot) neural network NEB studies such as CatTsunami~\cite{wander2025cattsunami}. The inputs to the network are the positions of the atoms, $x_i$, and the outputs are the forces on each atom and the potential energy of the system.

\subsection*{Lennard-Jones Transitions}\label{subsec1}
\begin{figure}[t]
\centering
\includegraphics[width=1.0\textwidth]{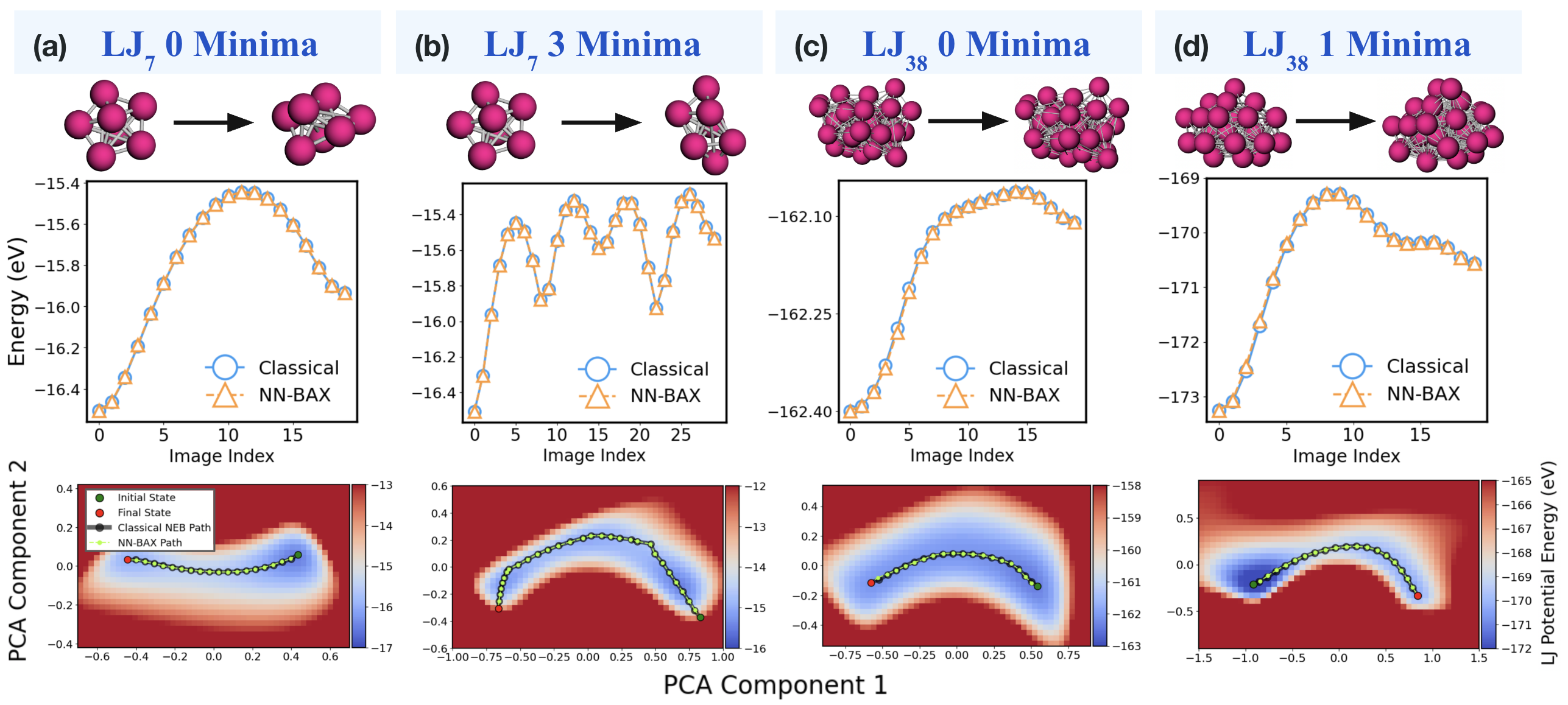}
\caption[baxbarriers]{Results from running NN-BAX on various paths with $\mathrm{LJ}_7$ and $\mathrm{LJ}_{38}$ systems. The upper row compares the energy profile predicted by NN-BAX to classical NEB, along with atomic visuals for the transition. The bottom row shows dimension-reduced principal component analysis plots of the energy, with the NN-BAX and classical NEB paths overlaid.
}
\label{fig:2}
\end{figure}

\begin{figure}
\centering
\includegraphics[width=1.0\textwidth]{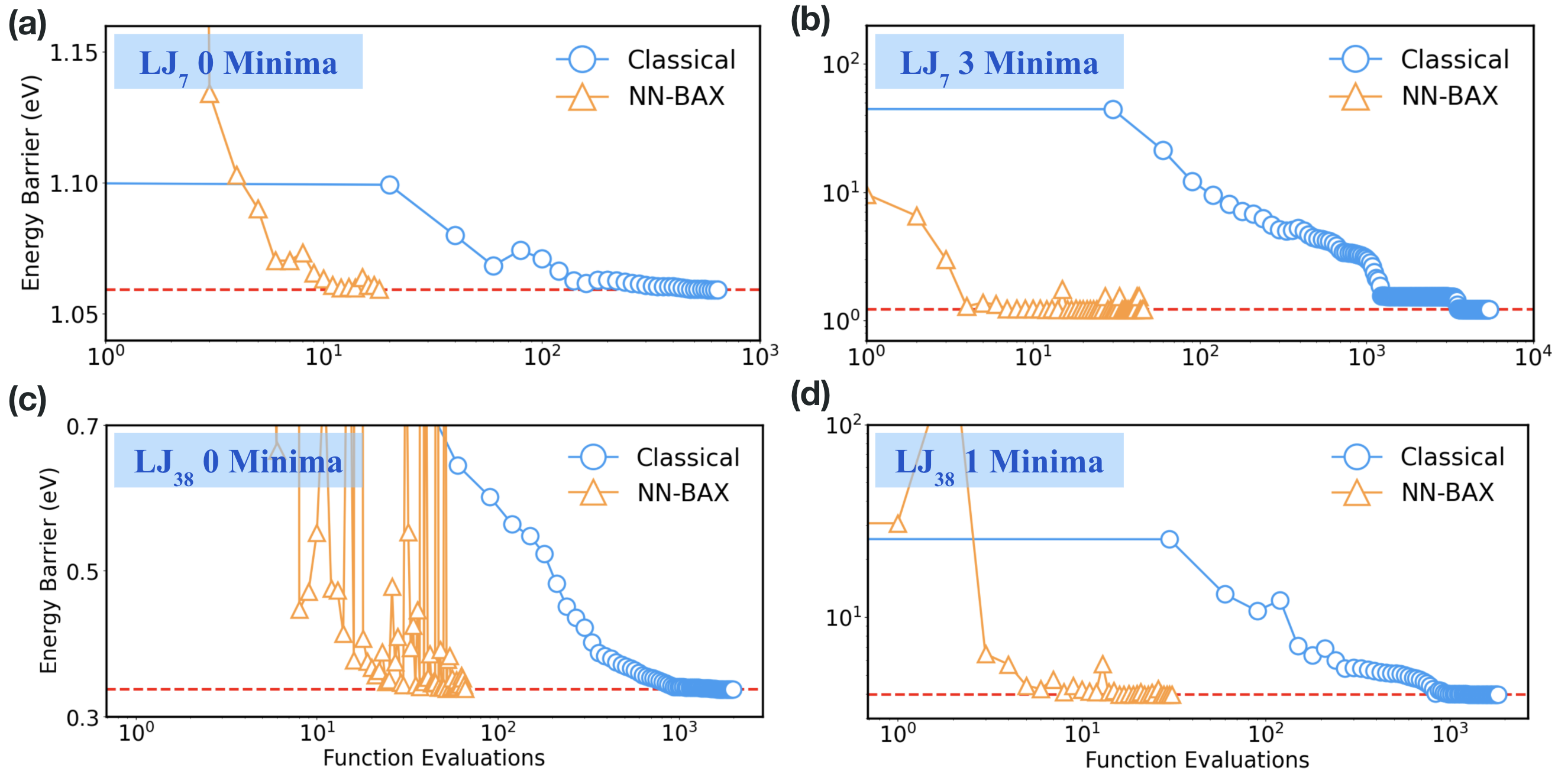}
\caption[Speedup]{Energy barrier v. function evaluations, for various $LJ_7$ and $LJ_{38}$ systems. Speedup factors are 34, 116, 20, and 38 for a), b), c), and d) respectively. The dashed red line denotes the energy barrier achieved by classical NEB. Note that for the NN-BAX curve, each point corresponds to an NEB optimization. We show here how quickly our method matches the ground truth for the energy barrier, the primary physical quantity of interest.
}
\label{fig:3}
\end{figure}

We demonstrate NN-BAX on atomic systems governed by the Lennard-Jones (LJ) potential:

\begin{equation}
V_{\mathrm{LJ}}(r) = 4\varepsilon \left[ \left(\frac{\sigma}{r}\right)^{12} - \left(\frac{\sigma}{r}\right)^{6} \right],
\end{equation}

because of its physical relevance and its well-studied transitions~\cite{wales1997global, doye2002saddle, schwerdtfeger2024100, wales1994rearrangements, doye1998thermodynamics, doye1999double, sehgal2014phase, wales1998archetypal, bohner2013quadratically}. Unlike energy/force computations involving density functional theory, the LJ potential is inexpensive to call. However, it retains much of the complexity of similar chemical systems, making it ideal for testing our method. We analyze systems with 7 and 38 identical atoms, abbreviated as $LJ_7$ and $LJ_{38}$. The dimensionality of these systems, particularly the 114-dimensional $LJ_{38}$, makes Gaussian process surrogates infeasible. We specifically choose N = 7, 38 for our $LJ_N$ systems because these systems exhibit many interesting local minima configurations, thus leading to them also being well-studied. $LJ_{38}$ is especially interesting because the two lowest energy configurations are well separated in atomic configuration space, meaning the transition between the two is nontrivial~\cite{trygubenko2004doubly}.

We analyze four transitions of varying difficulty, from $\mathrm{LJ}_7$ and $\mathrm{LJ}_{38}$. For $LJ_7$ we look at paths with 0 and 3 intermediate minima, and for $LJ_{38}$ we analyze paths with 0 and 1 intermediate minima. Paths with even more minima are incompatible with the NEB method used in this paper, and generally must be subdivided, as detailed in the Foundation-BAX section. These paths converged for classical NEB with $f_{max}$ values of 0.05, 0.15, 0.05, and 0.3\text{eV}/\text{\AA} respectively. We use the string method~\cite{weinan2002string}, which ignores spring forces and evenly redistributes images along the path every iteration. Figure~\ref{fig:2} compares the converged NN-BAX paths with their classical NEB counterparts. The visuals on top display the atomic structures of the initial and final states. The upper plots show energy versus image index for both classical NEB and NN-BAX, with very close agreement between the two results. Of the four, the worst predicted energy barrier is 0.3\% greater than the ground truth. The bottom row shows a two-dimensional principal component analysis (PCA) projection of the MEPs and potential energy landscapes, with projection computed from the ground truth path in atomic coordinate space. We observe that the NN-BAX paths qualitatively line up with the classical NEB paths, indicating that NN-BAX is finding the same structures as classical NEB.

Figure~\ref{fig:3} displays the speed-up NN-BAX achieves with respect to classical NEB, in terms of function evaluations. For systems running expensive DFT simulations, we expect the NN-BAX procedural overhead to be negligible, making number of function calls an appropriate figure of merit. We achieve a 1-2 orders of magnitude reduction in function calls across all four paths. Finally, we observe that the $LJ_7$ path with 3 minima achieves a greater speedup than the other paths. We hypothesize this occurs because while classical NEB time is determined by the complexity of NEB convergence (i.e. number of NEB iterations), NN-BAX is limited by modeling complexity of the potential energy surface. In NN-BAX, the convergence complexity has negligible impact because the NEB loop only calls the surrogate, and in the classical case this path requires $\sim$3x as many NEB iterations compared to the others. Thus, because the number of function evaluations in NN-BAX no longer scales with the number of NEB iterations, we observe a greater speedup for this path. We emphasize that NN-BAX has the greatest impact in cases where NEB complexity, specifically the number of NEB iterations required for convergence, is large. 

With the current NN-BAX implementation, we estimate the NN-BAX overhead to be equivalent to running classical NEB with simulations that each take approximately 25.4 seconds. For simulation times $\gg$25.4 seconds, as may be expected for DFT calculations on complex systems, the NN-BAX overhead becomes negligible. Details of this estimate, as well as suggestions of implementation optimizations to reduce the NN-BAX overhead, are given in the supplemental material.

\subsection*{Embedded Atom Method Potentials}\label{subsec2}

\begin{figure}
\centering
\includegraphics[width=1.0\textwidth]{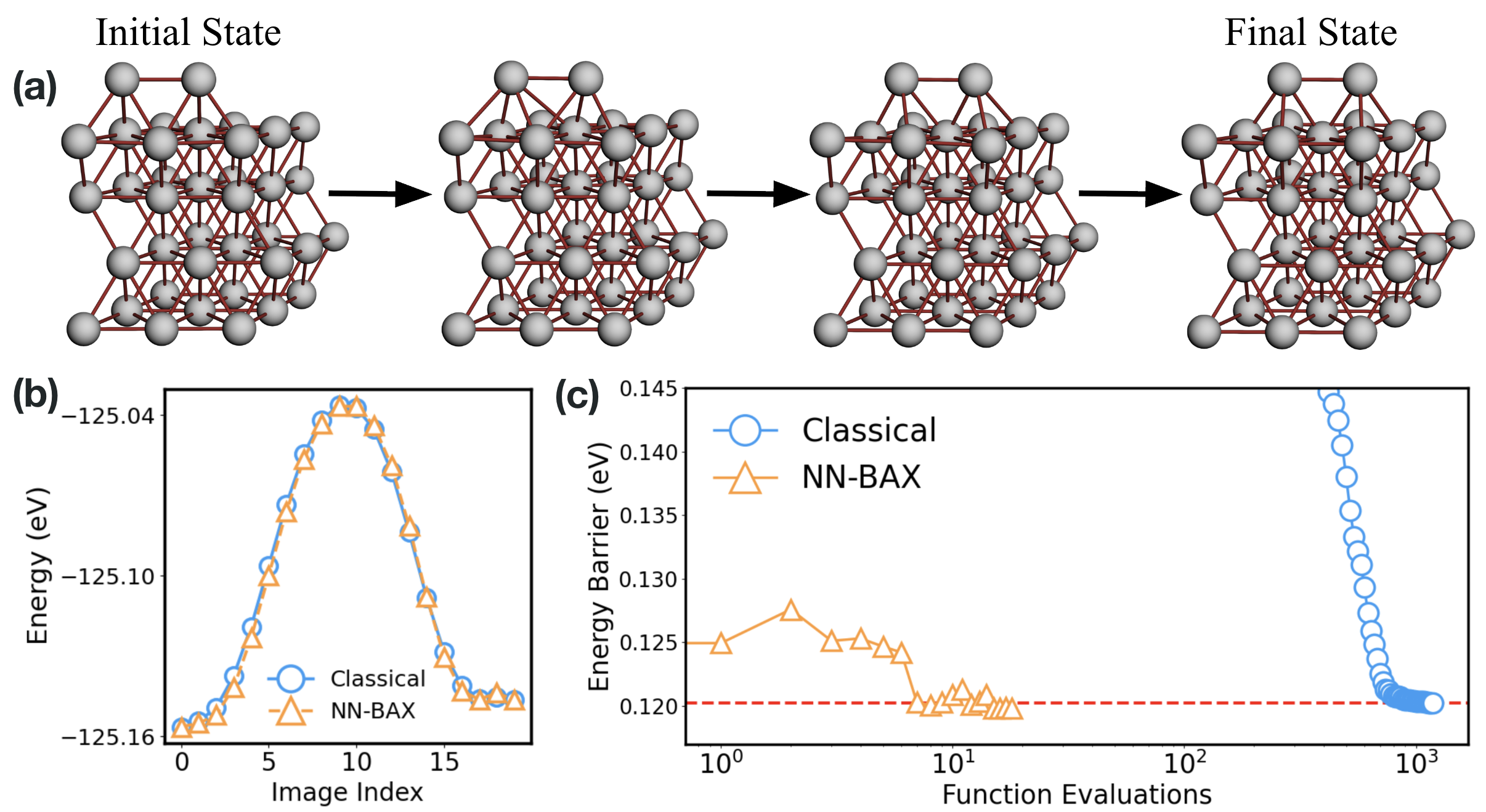}
\caption[copperFig]{ Embedded Atom Method potential with NN-BAX. a) displays a visual of the copper dimer diffusion transition. b) shows the energy profile from classical NEB and from NN-BAX. c) shows the energy barrier of classical NEB and NN-BAX, as a function of evaluations. The red line denotes the energy barrier achieved by classical NEB.
}
\label{fig:6} 
\end{figure}

Although Lennard-Jones structures and transitions host much complexity as a test system, the LJ potential is still pairwise, and thus does not capture some of the many-body nuance seen in DFT. Next, we test NN-BAX on an Embedded Atom Method (EAM) potential. EAM potentials are semi-empirical and designed especially for metals. They take into account both pairwise interactions and the collective electron density from neighboring atoms~\cite{daw1984embedded, foiles1986embedded}. This many-body formulation captures the physics of metals, for example surface diffusion~\cite{marinica2005diffusion, morgenstern2004direct, sbiaai2014diffusion, minkowski2015diffusion}. Surface diffusion plays a crucial role in processes such as thin-film growth, nanostructure formation, and nanoscale stability of metals~\cite{xia2013role, bonzel1990surface, wang2019surface, forgerini2014brief}.

We analyze a diffusion transition for copper involving a copper dimer moving to an adjacent site. Specifically, we analyze a 3x3x4 Cu(100) slab, with a pair of copper atoms atop. The bottom layer, consisting of nine atoms, is fixed to represent the bulk, and the rest of the atoms are free to move. Thus a total of 29 atoms are dynamic, giving us an 87 dimensional system. Note that while only the two atoms comprising the dimer differ in the initial and final states, the atoms beneath do move non-trivially during the transition. A visual for the transition is shown in Figure~\ref{fig:6}a. We use the potential described in~\cite{mishin2001structural}, which is specific to copper. For this transition we use the traditional NEB method with spring forces with \( k = 0.1\, \text{eV/\AA}^2 \), and observe that the path converges for $f_{max} < 0.01$\text{eV}/\text{\AA}. Shown in Figure~\ref{fig:6}b is the classical NEB energy profile and the NN-BAX energy profile. The two profiles match well, indicating that NN-BAX has found the correct path. Finally, Figure~\ref{fig:6}c shows the energy barrier versus function evaluations, and we observe a significant speedup from NN-BAX. Note that the initial energy barrier for the EAM potential is relatively better than those in the LJ systems because this system more closely resembles those in the OMat24 dataset.

\subsection*{Multi-step Transitions with Foundation-BAX}\label{subsec3}

\begin{figure}[t]
\centering
\includegraphics[width=1.0\textwidth]{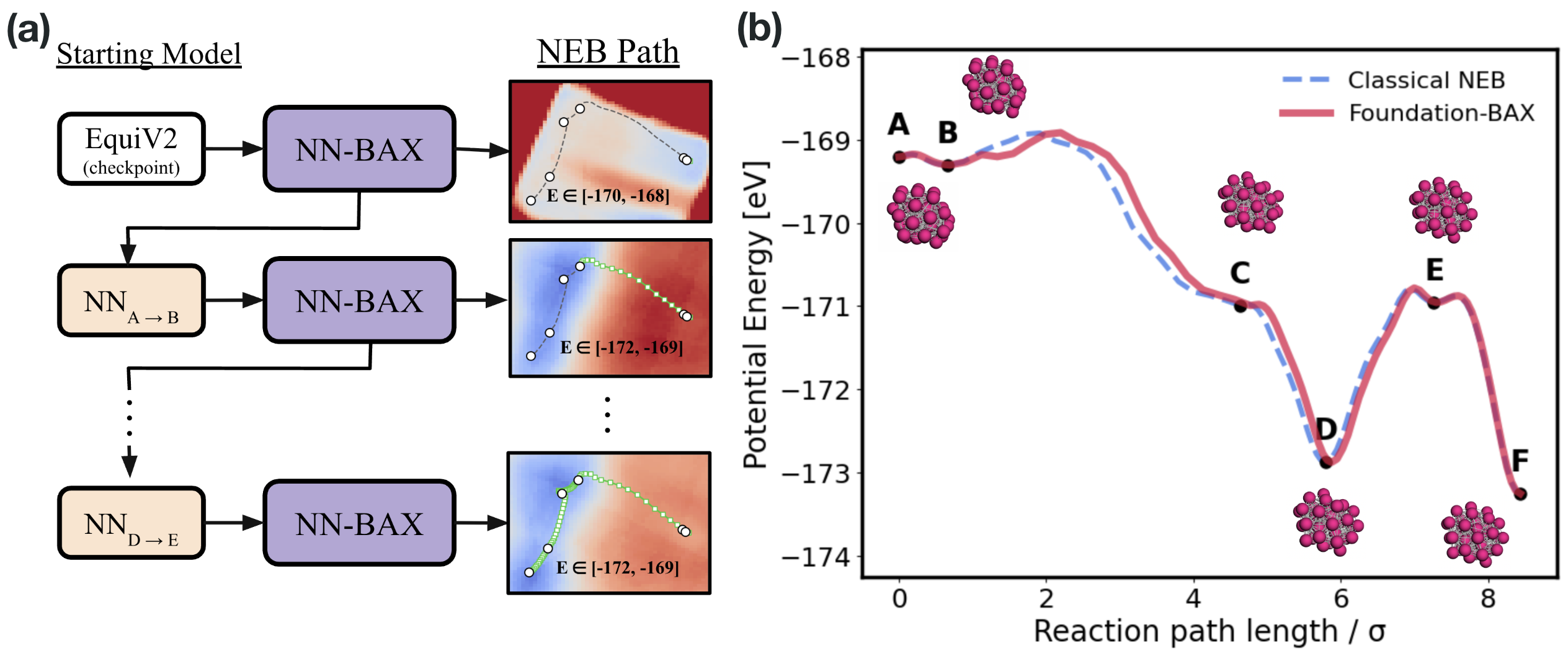}
\caption[Fbax]{ Foundation-BAX pipeline and path. a) displays the Foundation-BAX pipeline, where the model from the previous path is used as the starting model for the next path b) shows the $LJ_{38}$ path, which consists of 5 sub-transitions, connecting 6 local minima. 
}
\label{fig:4}
\end{figure}

For more complex transitions with many intermediate minima, the NEB algorithm often cannot converge to the correct path. Thus, it is common to use techniques to segment the transition into smaller paths between neighboring local minima. For Lennard-Jones transitions, the \texttt{OPTIM} program~\cite{wales2003optim} is used to search for complex multi-step transitions in high dimensional systems such as  $LJ_{38}$. In order to achieve an even greater speedup with respect to NN-BAX for such multi-step transitions, we introduce Foundation-BAX; i.e. we use BAX to assemble and train a foundation model on a shared pool of relevant simulations. One can imagine this technique applied broadly, where researchers who run algorithms on similar systems can collectively pool data to achieve more accurate surrogate models. In our case of multi-step transitions, implementing Foundation-BAX involves using information from previous paths to speed up NN-BAX for subsequent paths. Figure~\ref{fig:4}a displays the Foundation-BAX pipeline, where we use the final fine-tuned model of the previous path as the initial model for the neighboring path. 

To test the Foundation-BAX method, we choose a multi-step $LJ_{38}$ transition that connects six local minima, thus containing four intermediate minima. This transition specifically is a subset of a $LJ_{38}$ global minimum to second global minimum transition pathway, taken from example in~\cite{walesgroup_examples}. Figure~\ref{fig:4}b displays the potential energy of the true transition pathway, along with the potential energy of the path predicted by Foundation-BAX. We note that the overall shape and energy barrier of each transition are quite similar. Note that there is a slight shift in the predicted energy profile with respect to the true transition, particularly in the transition from B to C. This is a result of Foundation-BAX finding an equally energetically favorable yet marginally different path. 

\begin{figure}
\centering
\includegraphics[width=1.0\textwidth]{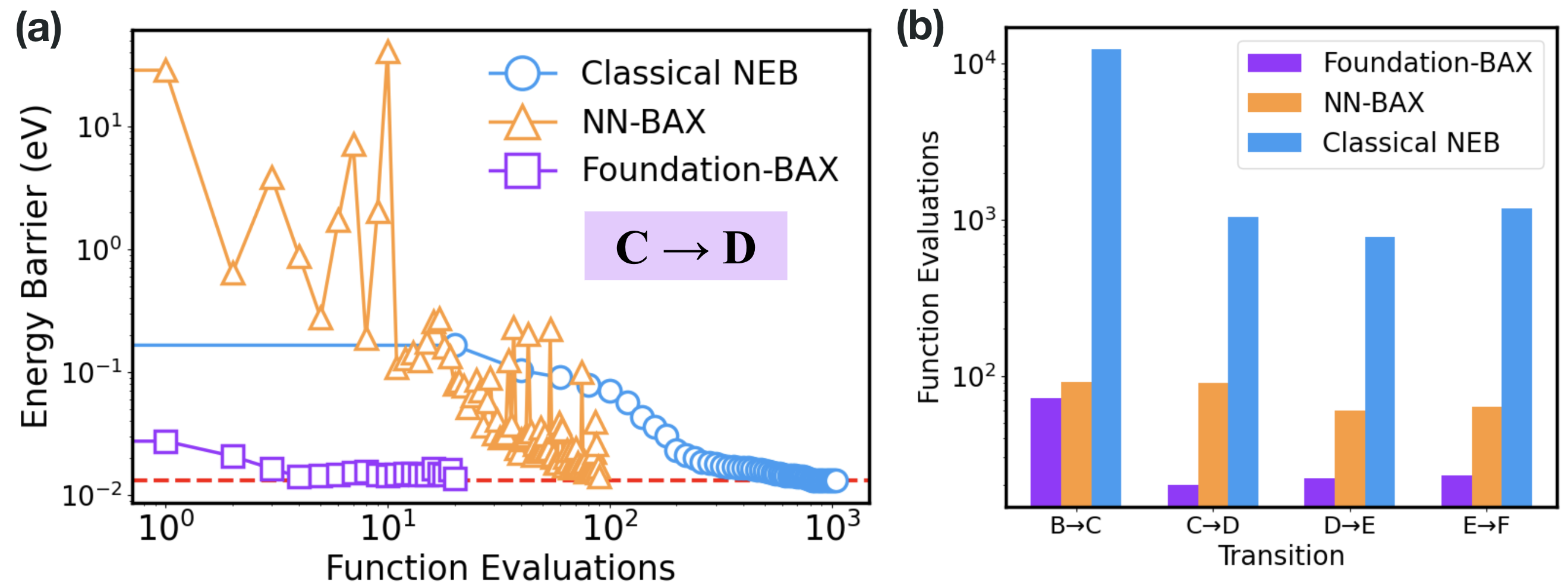}
\caption[fbaxSpeedup]{ Foundation-BAX speedup. a) Compares the number of function evaluations for Foundation-BAX, NN-BAX, and Classical NEB for the C to D transition. b) Shows the total number of function evaluations required for all three methods for all paths.
}
\label{fig:5}
\end{figure}

Figure~\ref{fig:5} displays the speedup, in terms of function evaluations, achieved by Foundation-BAX, compared to classical NEB and NN-BAX. Figure~\ref{fig:5}a displays the energy barrier v. function evaluation for Foundation-BAX, NN-BAX, and classical NEB, for the C to D transition. We observe that Foundation-BAX has a 4.5$\times$ speedup compared to NN-BAX and a 52$\times$ speedup compared to classical NEB. Note the significantly flatter nature of the Foundation-BAX curve, implying that some of the force landscape has already been learned, leading to a quicker convergence. Figure~\ref{fig:5}b compares the total number of function evaluations for all three methods, for all paths. We start with a standard NN-BAX run for the A to B transition. For subsequent runs, we observe that Foundation-BAX requires fewer function evaluations to converge compared to NN-BAX. Furthermore, the speedup with respect to NN-BAX for the final three transitions are greater than that of the B to C transition, indicating that the model benefited by learning from more paths.  Finally we note that Foundation-BAX works optimally with different convergence hyperparameters than NN-BAX, and elaborate more in supplemental materials. 

\section*{Discussion}\label{sec3}

Accurately characterizing the transformations between metastable states is key to understanding and predicting the behavior of materials and chemical systems. The nudged elastic band algorithm is critical in finding transition states and energy barriers in such chemical systems. Systems modeled by Lennard-Jones or Embedded Atom Method potentials reproduce key features of complex chemical interactions, providing a controlled framework for studying high-dimensional atomic dynamics. By applying our methods to these high dimensional systems, we unify efforts in active learning and foundation models using BAX. Specifically, we apply NN-BAX to NEB on LJ and EAM potentials and observe a speedup of 1-2 orders of magnitude in terms of number of function evaluations. Furthermore, we observe that it is possible to gain a greater advantage in multi-step transitions with Foundation-BAX, where we fine-tune on previous fine-tuned models. We anticipate that Foundation-BAX can extend beyond multi-step transitions and be used more broadly to accelerate the speed of a suite of jobs governed by the same potential.

Previous use of BAX has been restricted to Gaussian process surrogates, which famously suffer from the curse of dimensionality~\cite{binois2022survey}. The primary advantage of NN-BAX relies on the introduction of a neural network as a surrogate model in the BAX procedure, which allows for scalability to higher dimensional systems. We emphasize that while LJ and EAM do not require expensive \textit{ab initio} calculations, our work with NN-BAX describes a broader system-agnostic procedure. Therefore, we expect our work to extend to both new surrogate models and NEB applications, including high-dimensional systems that require expensive simulations like DFT. This technique may help us understand systems currently infeasible to study, including complex catalytic reactions and conformational pathways in proteins~\cite{wander2025cattsunami, mathews2006nudged}. Estimating 1000 \mbox{energy/force} evaluations per NEB optimization in~\cite{wander2025cattsunami}, we estimate $\sim$14.4 minutes per DFT calculation for catalysts. This fits the $\gg$25.4 seconds we present above in order for the NN-BAX overhead to be negligible. With some improvements to our implementation (see supplementary material) we expect that we can reduce the NN-BAX overhead to $\sim$3.1 seconds, further amplifying applicability.

This work shows that coupling algorithm-aware active learning with expressive, symmetry-aware neural network surrogates can substantially reduce the cost of transition state discovery. Our approach provides a path towards expediting high-dimensional transition-state calculations for complex systems in physics, chemistry, and biology, including those which are currently computationally intractable.

\section*{Methods}\label{sec4}

\subsection*{NN-BAX Algorithm}\label{sec4subsec1}
There are a variety of forms of BAX acquisition. InfoBAX~\cite{neiswanger2021bayesian}, an information-based acquisition function, aims to maximize the expected information gain (EIG) about an algorithm output. However, it is costly to compute, involving several model retrainings to estimate the associated entropies. Heuristic or approximate BAX algorithms have been proposed that are much more efficient, such as MeanBAX~\cite{chitturi2024targeted} and PS-BAX~\cite{cheng2024practical} which run the algorithm on the posterior mean or a posterior sample and then take the output point with the highest predicted uncertainty, avoiding the cost of computing the full EIG. In this work, we take a similar, heuristic approach. Our BAX acquisition involves running NEB on a deterministic model of the potential energy/forces. We then randomly sample from the resulting NEB path to choose the next point to acquire and update the model for the next loop. Our design also leaves open a clear path toward incorporating more rigorous acquisition strategies that involve uncertainty quantification, such as MeanBAX and InfoBAX, which may yield additional gains. 

\subsection*{System Setup and Training}\label{sec4subsec2}
To create a baseline, we first find local minima using basin hopping~\cite{wales1997global}. We then run classical NEB between pairs of local minima, varying $f_{max}$ as needed to achieve convergence. To check stability of the resulting MEPs, we randomly perturb the positions of the images from the converged path by $\sigma_x \in [0,\, 0.1]~\text{\AA}$ and re-run the NEB minimization. If after 100 iterations the path does not move and no lower energy path is discovered, the MEP is considered a ``ground truth'' target for our NN-BAX optimization. For both classical and NN-BAX NEB calculation, we use the Atomic Simulation Environment (ASE) library~\cite{larsen2017atomic}. All NEB runs in this paper use the FIRE optimizer~\cite{bitzek2006structural} with parameters $dt = 0.1$ and $dt_{max} = 1.0$. 

One problem with running NEB on a surrogate is that for less accurate models in earlier iterations, NEB often does not converge below the given $f_{max}$ criterion. Thus we must set some finite number of NEB iterations, $N_{neb}$ otherwise the process would never converge. However for NN-BAX, unlike classical NEB, the number of function evaluations does not scale with $N_{neb}$. Thus, we introduce a hyperparameter $N_{neb, max}$, where every BAX iteration runs NEB for $N_{neb, max}$ iterations, and the path with the lowest $f_{max}$ is selected. We find that $N_{neb, max} = 200$ works well for the initial suite of LJ transitions. For the EAM transition we use $N_{neb, max} = 100$. Finally, for the Foundation-BAX pathways, we used $N_{neb, max} = 100$ for all segments except from B to C, where we use a larger $N_{neb, max} = 650$.

In this paper we introduce a new convergence hyperparameter, $\mathrm{MAE}_i$, which corresponds to the mean absolute error (MAE) in the prediction of model $i-1$ on the acquired image in iteration $i$. Reduction in $\mathrm{MAE}_i$ indicates that the model is performing better on points not in the training set and, once NN-BAX begins predicting similar final paths, the sampled points grow increasingly similar, making for an easier modeling task. For NN-BAX to converge, $\mathrm{MAE}_i < m$ and $f_{max, BAX}^i < t$ must be satisfied simultaneously, where $m$ and $t$ are the respective convergence thresholds. For $\mathrm{MAE}_i$ convergence, $m = 0.1$ proved to generally work. For force convergence, we set
$f_{\mathrm{max},\mathrm{BAX}} = n \, f_{\mathrm{max},\mathrm{classical}}$, 
where $n > 1$ accounts for small amounts of noise in the modeled predictions with respect to true potential. In practice, we found $n = 2$ to work well for our paths. For further confidence in convergence we introduce a patience metric $p$, where the convergence criteria must be achieved $p$ times for NN-BAX to converge. $p=2$ was found to be generally robust. 

In each BAX iteration, the network was trained for 50 epochs with a batch size of 2 using the AdamW optimizer~\cite{loshchilov2017decoupled} (weight decay $0.001$, gradient clipping at $100$) and an initial learning rate of $2\times 10^{-4}$. A cosine learning rate schedule was employed, with a warmup phase lasting $1$~epoch (warmup factor $0.2$) and a minimum learning rate set to $1\%$ of the initial value. Exponential moving average (EMA) updates with a decay of $0.999$ were applied throughout training. We use the mean absolute error (MAE) loss, giving forces four times the weight of energy since our convergence criterion is force-based. 
All training hyperparameters were the same those fairchem used to train the EquiformerV2 checkpoint~\cite{facebookresearch_fairchem}, except for batch size and epochs which were modified to accommodate smaller dataset sizes. All models are trained on a single NVIDIA Tesla A100 GPU at the SLAC Shared Science Data Facility.

\backmatter

\bmhead{Acknowledgements}
We acknowledge Guanzhi Li, Yu Lin, Minkyung Han, Yu Zhang, Samuel Klein, and David Wales for helpful insights and manuscript feedback. SG and PK were supported by the Department of Energy, Laboratory Directed Research and Development program at SLAC National Accelerator Laboratory, under contract DE-AC02-76SF00515.

\bmhead{Code availability}
We provide a user-friendly implementation of NN-BAX with the nudged elastic band algorithm at: \url{https://github.com/pranavkakhandiki/nn-bax-neb}. This repository also contains example notebooks to aid in analysis. 

\bmhead{Competing interests}
The authors declare no competing interests.

\bibliography{sn-bibliography}

@article{wander2025cattsunami,
  title={CatTSunami: Accelerating transition state energy calculations with pretrained graph neural networks},
  author={Wander, Brook and Shuaibi, Muhammed and Kitchin, John R and Ulissi, Zachary W and Zitnick, C Lawrence},
  journal={ACS Catalysis},
  volume={15},
  number={7},
  pages={5283--5294},
  year={2025},
  publisher={ACS Publications}
}

@article{garrido2019low,
  title={Low-scaling algorithm for nudged elastic band calculations using a surrogate machine learning model},
  author={Garrido Torres, Jos{\'e} A and Jennings, Paul C and Hansen, Martin H and Boes, Jacob R and Bligaard, Thomas},
  journal={Physical review letters},
  volume={122},
  number={15},
  pages={156001},
  year={2019},
  publisher={APS}
}

@article{koistinen2019nudged,
  title={Nudged elastic band calculations accelerated with Gaussian process regression based on inverse interatomic distances},
  author={Koistinen, Olli-Pekka and {\'A}sgeirsson, Vilhj{\'a}lmur and Vehtari, Aki and J{\'o}nsson, Hannes},
  journal={Journal of chemical theory and computation},
  volume={15},
  number={12},
  pages={6738--6751},
  year={2019},
  publisher={ACS Publications}
}

@article{fu2025learning,
  title={Learning smooth and expressive interatomic potentials for physical property prediction},
  author={Fu, Xiang and Wood, Brandon M and Barroso-Luque, Luis and Levine, Daniel S and Gao, Meng and Dzamba, Misko and Zitnick, C Lawrence},
  journal={arXiv preprint arXiv:2502.12147},
  year={2025}
}

@article{luo2025automatic,
  title={Automatic identification of slip pathways in ductile inorganic materials by combining the active learning strategy and NEB method},
  author={Luo, Jun and Fan, Tao and Zhang, Jiawei and Qiu, Pengfei and Shi, Xun and Chen, Lidong},
  journal={npj Computational Materials},
  volume={11},
  number={1},
  pages={41},
  year={2025},
  publisher={Nature Publishing Group UK London}
}

@article{deng2025systematic,
  title={Systematic softening in universal machine learning interatomic potentials},
  author={Deng, Bowen and Choi, Yunyeong and Zhong, Peichen and Riebesell, Janosh and Anand, Shashwat and Li, Zhuohan and Jun, KyuJung and Persson, Kristin A and Ceder, Gerbrand},
  journal={npj Computational Materials},
  volume={11},
  number={1},
  pages={9},
  year={2025},
  publisher={Nature Publishing Group UK London}
}

@article{li2025transition,
  title={Transition state searching accelerated by neural network potential},
  author={Li, Bowen and Xiao, Jin and Gao, Ya and Zhang, John ZH and Zhu, Tong},
  journal={Journal of Chemical Information and Modeling},
  volume={65},
  number={5},
  pages={2297--2303},
  year={2025},
  publisher={ACS Publications}
}

@article{perego2024data,
  title={Data efficient machine learning potentials for modeling catalytic reactivity via active learning and enhanced sampling},
  author={Perego, Simone and Bonati, Luigi},
  journal={npj Computational Materials},
  volume={10},
  number={1},
  pages={291},
  year={2024},
  publisher={Nature Publishing Group UK London}
}

@article{schaaf2023accurate,
  title={Accurate energy barriers for catalytic reaction pathways: an automatic training protocol for machine learning force fields},
  author={Schaaf, Lars L and Fako, Edvin and De, Sandip and Sch{\"a}fer, Ansgar and Cs{\'a}nyi, G{\'a}bor},
  journal={npj Computational Materials},
  volume={9},
  number={1},
  pages={180},
  year={2023},
  publisher={Nature Publishing Group UK London}
}

@article{meyer2019machine,
  title={Machine learning in computational chemistry: An evaluation of method performance for nudged elastic band calculations},
  author={Meyer, Ralf and Schmuck, Klemens S and Hauser, Andreas W},
  journal={Journal of Chemical Theory and Computation},
  volume={15},
  number={11},
  pages={6513--6523},
  year={2019},
  publisher={ACS Publications}
}

@article{koistinen2017nudged,
  title={Nudged elastic band calculations accelerated with Gaussian process regression},
  author={Koistinen, Olli-Pekka and Dagbjartsd{\'o}ttir, Freyja B and {\'A}sgeirsson, Vilhj{\'a}lmur and Vehtari, Aki and J{\'o}nsson, Hannes},
  journal={The Journal of chemical physics},
  volume={147},
  number={15},
  year={2017},
  publisher={AIP Publishing}
}

@article{schreiner2022neuralneb,
  title={NeuralNEB—neural networks can find reaction paths fast},
  author={Schreiner, Mathias and Bhowmik, Arghya and Vegge, Tejs and J{\o}rgensen, Peter Bj{\o}rn and Winther, Ole},
  journal={Machine Learning: Science and Technology},
  volume={3},
  number={4},
  pages={045022},
  year={2022},
  publisher={IOP Publishing}
}

@article{teng2024physical,
  title={Physical prior mean function-driven Gaussian processes search for minimum-energy reaction paths with a climbing-image nudged elastic band: a general method for gas-phase, interfacial, and bulk-phase reactions},
  author={Teng, Chong and Wang, Yang and Bao, Junwei Lucas},
  journal={Journal of Chemical Theory and Computation},
  volume={20},
  number={10},
  pages={4308--4324},
  year={2024},
  publisher={ACS Publications}
}

@inproceedings{neiswanger2021bayesian,
  title={Bayesian algorithm execution: Estimating computable properties of black-box functions using mutual information},
  author={Neiswanger, Willie and Wang, Ke Alexander and Ermon, Stefano},
  booktitle={International Conference on Machine Learning},
  pages={8005--8015},
  year={2021},
  organization={PMLR}
}

@article{chitturi2024targeted,
  title={Targeted materials discovery using Bayesian algorithm execution},
  author={Chitturi, Sathya R and Ramdas, Akash and Wu, Yue and Rohr, Brian and Ermon, Stefano and Dionne, Jennifer and Jornada, Felipe H da and Dunne, Mike and Tassone, Christopher and Neiswanger, Willie and others},
  journal={npj Computational Materials},
  volume={10},
  number={1},
  pages={156},
  year={2024},
  publisher={Nature Publishing Group UK London}
}

@incollection{jonsson1998nudged,
  title={Nudged elastic band method for finding minimum energy paths of transitions},
  author={J{\'o}nsson, Hannes and Mills, Greg and Jacobsen, Karsten W},
  booktitle={Classical and quantum dynamics in condensed phase simulations},
  pages={385--404},
  year={1998},
  publisher={World Scientific}
}

@article{henkelman2000improved,
  title={Improved tangent estimate in the nudged elastic band method for finding minimum energy paths and saddle points},
  author={Henkelman, Graeme and J{\'o}nsson, Hannes},
  journal={The Journal of chemical physics},
  volume={113},
  number={22},
  pages={9978--9985},
  year={2000},
  publisher={American Institute of Physics}
}

@article{henkelman2000climbing,
  title={A climbing image nudged elastic band method for finding saddle points and minimum energy paths},
  author={Henkelman, Graeme and Uberuaga, Blas P and J{\'o}nsson, Hannes},
  journal={The Journal of chemical physics},
  volume={113},
  number={22},
  pages={9901--9904},
  year={2000},
  publisher={American Institute of Physics}
}

@article{trygubenko2004doubly,
  title={A doubly nudged elastic band method for finding transition states},
  author={Trygubenko, Semen A and Wales, David J},
  journal={The Journal of chemical physics},
  volume={120},
  number={5},
  pages={2082--2094},
  year={2004},
  publisher={AIP Publishing}
}

@article{weinan2002string,
  title={String method for the study of rare events},
  author={Weinan, E and Ren, Weiqing and Vanden-Eijnden, Eric},
  journal={Physical Review B},
  volume={66},
  number={5},
  pages={052301},
  year={2002},
  publisher={APS}
}

@article{maragakis2002adaptive,
  title={Adaptive nudged elastic band approach for transition state calculation},
  author={Maragakis, P and Andreev, Stefan A and Brumer, Yisroel and Reichman, David R and Kaxiras, Efthimios},
  journal={The Journal of chemical physics},
  volume={117},
  number={10},
  pages={4651--4658},
  year={2002},
  publisher={American Institute of Physics}
}

@article{sheppard2012generalized,
  title={A generalized solid-state nudged elastic band method},
  author={Sheppard, Daniel and Xiao, Penghao and Chemelewski, William and Johnson, Duane D and Henkelman, Graeme},
  journal={The Journal of chemical physics},
  volume={136},
  number={7},
  year={2012},
  publisher={AIP Publishing}
}

@article{kolsbjerg2016automated,
  title={An automated nudged elastic band method},
  author={Kolsbjerg, Esben L and Groves, Michael N and Hammer, Bj{\o}rk},
  journal={The Journal of chemical physics},
  volume={145},
  number={9},
  year={2016},
  publisher={AIP Publishing}
}

@article{wales1997global,
  title={Global optimization by basin-hopping and the lowest energy structures of Lennard-Jones clusters containing up to 110 atoms},
  author={Wales, David J and Doye, Jonathan PK},
  journal={The Journal of Physical Chemistry A},
  volume={101},
  number={28},
  pages={5111--5116},
  year={1997},
  publisher={ACS Publications}
}

@article{doye2002saddle,
  title={Saddle points and dynamics of Lennard-Jones clusters, solids, and supercooled liquids},
  author={Doye, Jonathan PK and Wales, David J},
  journal={The Journal of Chemical Physics},
  volume={116},
  number={9},
  pages={3777--3788},
  year={2002},
  publisher={American Institute of Physics}
}

@article{schwerdtfeger2024100,
  title={100 years of the Lennard-Jones potential},
  author={Schwerdtfeger, Peter and Wales, David J},
  journal={Journal of Chemical Theory and Computation},
  volume={20},
  number={9},
  pages={3379--3405},
  year={2024},
  publisher={ACS Publications}
}

@article{wales1994rearrangements,
  title={Rearrangements of 55-atom Lennard-Jones and (C60) 55 clusters},
  author={Wales, David J},
  journal={The Journal of chemical physics},
  volume={101},
  number={5},
  pages={3750--3762},
  year={1994},
  publisher={American Institute of Physics}
}

@article{doye1998thermodynamics,
  title={Thermodynamics and the global optimization of Lennard-Jones clusters},
  author={Doye, Jonathan PK and Wales, David J and Miller, Mark A},
  journal={The Journal of Chemical Physics},
  volume={109},
  number={19},
  pages={8143--8153},
  year={1998},
  publisher={American Institute of Physics}
}

@article{doye1999double,
  title={The double-funnel energy landscape of the 38-atom Lennard-Jones cluster},
  author={Doye, Jonathan PK and Miller, Mark A and Wales, David J},
  journal={The Journal of Chemical Physics},
  volume={110},
  number={14},
  pages={6896--6906},
  year={1999},
  publisher={American Institute of Physics}
}

@article{sehgal2014phase,
  title={Phase behavior of the 38-atom Lennard-Jones cluster},
  author={Sehgal, Ray M and Maroudas, Dimitrios and Ford, David M},
  journal={The Journal of Chemical Physics},
  volume={140},
  number={10},
  year={2014},
  publisher={AIP Publishing}
}

@article{wales1998archetypal,
  title={Archetypal energy landscapes},
  author={Wales, David J and Miller, Mark A and Walsh, Tiffany R},
  journal={Nature},
  volume={394},
  number={6695},
  pages={758--760},
  year={1998},
  publisher={Nature Publishing Group UK London}
}

@article{bohner2013quadratically,
  title={A quadratically-converging nudged elastic band optimizer},
  author={Bohner, Matthias U and Meisner, Jan and K{\"a}stner, Johannes},
  journal={Journal of Chemical Theory and Computation},
  volume={9},
  number={8},
  pages={3498--3504},
  year={2013},
  publisher={ACS Publications}
}

@article{liao2023equiformerv2,
  title={Equiformerv2: Improved equivariant transformer for scaling to higher-degree representations},
  author={Liao, Yi-Lun and Wood, Brandon and Das, Abhishek and Smidt, Tess},
  journal={arXiv preprint arXiv:2306.12059},
  year={2023}
}

@article{vaswani2017attention,
  title={Attention is all you need},
  author={Vaswani, Ashish and Shazeer, Noam and Parmar, Niki and Uszkoreit, Jakob and Jones, Llion and Gomez, Aidan N and Kaiser, {\L}ukasz and Polosukhin, Illia},
  journal={Advances in neural information processing systems},
  volume={30},
  year={2017}
}

@article{barroso2024open,
  title={Open materials 2024 (omat24) inorganic materials dataset and models},
  author={Barroso-Luque, Luis and Shuaibi, Muhammed and Fu, Xiang and Wood, Brandon M and Dzamba, Misko and Gao, Meng and Rizvi, Ammar and Zitnick, C Lawrence and Ulissi, Zachary W},
  journal={arXiv preprint arXiv:2410.12771},
  year={2024}
}

@article{bitzek2006structural,
  title={Structural relaxation made simple},
  author={Bitzek, Erik and Koskinen, Pekka and G{\"a}hler, Franz and Moseler, Michael and Gumbsch, Peter},
  journal={Physical review letters},
  volume={97},
  number={17},
  pages={170201},
  year={2006},
  publisher={APS}
}

@article{larsen2017atomic,
  title={The atomic simulation environment—a Python library for working with atoms},
  author={Larsen, Ask Hjorth and Mortensen, Jens J{\o}rgen and Blomqvist, Jakob and Castelli, Ivano E and Christensen, Rune and Du{\l}ak, Marcin and Friis, Jesper and Groves, Michael N and Hammer, Bj{\o}rk and Hargus, Cory and others},
  journal={Journal of Physics: Condensed Matter},
  volume={29},
  number={27},
  pages={273002},
  year={2017},
  publisher={IOP Publishing}
}

@article{jones1998efficient,
  title={Efficient global optimization of expensive black-box functions},
  author={Jones, Donald R and Schonlau, Matthias and Welch, William J},
  journal={Journal of Global optimization},
  volume={13},
  number={4},
  pages={455--492},
  year={1998},
  publisher={Springer}
}

@article{loshchilov2017decoupled,
  title={Decoupled weight decay regularization},
  author={Loshchilov, Ilya and Hutter, Frank},
  journal={arXiv preprint arXiv:1711.05101},
  year={2017}
}

@article{cheng2024practical,
  title={Practical Bayesian algorithm execution via posterior sampling},
  author={Cheng, Chu Xin and Astudillo, Raul and Desautels, Thomas A and Yue, Yisong},
  journal={Advances in Neural Information Processing Systems},
  volume={37},
  pages={135186--135207},
  year={2024}
}

@article{eyring1935activated,
  title={The activated complex in chemical reactions},
  author={Eyring, Henry},
  journal={The Journal of chemical physics},
  volume={3},
  number={2},
  pages={107--115},
  year={1935},
  publisher={American Institute of Physics}
}

@article{riebesell2023matbench,
  title={Matbench Discovery--A framework to evaluate machine learning crystal stability predictions},
  author={Riebesell, Janosh and Goodall, Rhys EA and Benner, Philipp and Chiang, Yuan and Deng, Bowen and Lee, Alpha A and Jain, Anubhav and Persson, Kristin A},
  journal={arXiv preprint arXiv:2308.14920},
  year={2023}
}

@article{wales2003optim,
  title={OPTIM: A program for optimizing geometries and calculating reaction pathways},
  author={Wales, DJ},
  journal={URL http://www-wales. ch. cam. ac. uk/software. html},
  year={2003}
}

@software{walesgroup_examples,
  author       = {Wales Group},
  title        = {Examples from the Wales Group Repository},
  url          = {https://github.com/wales-group/examples},
  version      = {master},
  organization = {University of Cambridge},
  year         = {2024},
  note         = {Accessed: 2025-11-03}
}

@article{daw1984embedded,
  title={Embedded-atom method: Derivation and application to impurities, surfaces, and other defects in metals},
  author={Daw, Murray S and Baskes, Michael I},
  journal={Physical review B},
  volume={29},
  number={12},
  pages={6443},
  year={1984},
  publisher={APS}
}

@article{foiles1986embedded,
  title={Embedded-atom-method functions for the fcc metals Cu, Ag, Au, Ni, Pd, Pt, and their alloys},
  author={Foiles, SM and Baskes, MI and Daw, Murray S},
  journal={Physical review B},
  volume={33},
  number={12},
  pages={7983},
  year={1986},
  publisher={APS}
}

@article{marinica2005diffusion,
  title={Diffusion rates of Cu adatoms on Cu (111) in the presence of an adisland nucleated at fcc or hcp sites},
  author={Marinica, Mihai-Cosmin and Barreteau, Cyrille and Spanjaard, Daniel and Desjonqu{\`e}res, Marie-Catherine},
  journal={Physical Review B—Condensed Matter and Materials Physics},
  volume={72},
  number={11},
  pages={115402},
  year={2005},
  publisher={APS}
}

@article{morgenstern2004direct,
  title={Direct imaging of Cu dimer formation, motion, and interaction with Cu atoms on Ag (111)},
  author={Morgenstern, Karina and Braun, Kai-Felix and Rieder, Karl-Heinz},
  journal={Physical review letters},
  volume={93},
  number={5},
  pages={056102},
  year={2004},
  publisher={APS}
}

@article{sbiaai2014diffusion,
  title={Diffusion processes of trimers on missing row surfaces: Cu 3/Ag (110) and Ag 3/Cu (110)},
  author={Sbiaai, Khalid and Eddiai, Adil and Boughaleb, Yahia and Mazroui, M’hammed and Raty, Jean-Yves and Meddad, Mounir and Kara, Abdelkader},
  journal={Optical and Quantum Electronics},
  volume={46},
  number={1},
  pages={15--22},
  year={2014},
  publisher={Springer}
}

@article{minkowski2015diffusion,
  title={Diffusion of Cu adatoms and dimers on Cu (111) and Ag (111) surfaces},
  author={Mi{\'n}kowski, Marcin and Za{\l}uska-Kotur, Magdalena A},
  journal={Surface Science},
  volume={642},
  pages={22--32},
  year={2015},
  publisher={Elsevier}
}

@article{xia2013role,
  title={On the role of surface diffusion in determining the shape or morphology of noble-metal nanocrystals},
  author={Xia, Xiaohu and Xie, Shuifen and Liu, Maochang and Peng, Hsin-Chieh and Lu, Ning and Wang, Jinguo and Kim, Moon J and Xia, Younan},
  journal={Proceedings of the National Academy of Sciences},
  volume={110},
  number={17},
  pages={6669--6673},
  year={2013},
  publisher={National Academy of Sciences}
}

@inbook{bonzel1990surface,
	address = {Dordrecht},
	author = {Bonzel, H. P.},
	booktitle = {Diffusion in Materials},
	editor = {Laskar, A. L. and Bocquet, J. L. and Brebec, G. and Monty, C.},
	isbn = {978-94-009-1976-1},
	pages = {307--308},
	publisher = {Springer Netherlands},
	title = {Surface Diffusion on Metals},
	year = {1990}
}

@article{wang2019surface,
  title={Surface diffusion-limited lifetime of silver and copper nanofilaments in resistive switching devices},
  author={Wang, Wei and Wang, Ming and Ambrosi, Elia and Bricalli, Alessandro and Laudato, Mario and Sun, Zhong and Chen, Xiaodong and Ielmini, Daniele},
  journal={Nature communications},
  volume={10},
  number={1},
  pages={81},
  year={2019},
  publisher={Nature Publishing Group UK London}
}

@article{forgerini2014brief,
  title={A brief review of mathematical models of thin film growth and surfaces: a possible route to avoid defects in stents},
  author={Forgerini, Fabricio L and Marchiori, Roberto},
  journal={Biomatter},
  volume={4},
  number={1},
  pages={e28871},
  year={2014},
  publisher={Taylor \& Francis}
}

@article{mishin2001structural,
  title={Structural stability and lattice defects in copper: Ab initio, tight-binding, and embedded-atom calculations},
  author={Mishin, Yuri and Mehl, Michael J and Papaconstantopoulos, Dimitrios A and Voter, Arthur F and Kress, Joel D},
  journal={Physical Review B},
  volume={63},
  number={22},
  pages={224106},
  year={2001},
  publisher={APS}
}

@article{piskulich2019activation,
  title={Activation energies and beyond},
  author={Piskulich, Zeke A and Mesele, Oluwaseun O and Thompson, Ward H},
  journal={The Journal of Physical Chemistry A},
  volume={123},
  number={33},
  pages={7185--7194},
  year={2019},
  publisher={ACS Publications}
}

@article{binois2022survey,
  title={A survey on high-dimensional Gaussian process modeling with application to Bayesian optimization},
  author={Binois, Mickael and Wycoff, Nathan},
  journal={ACM Transactions on Evolutionary Learning and Optimization},
  volume={2},
  number={2},
  pages={1--26},
  year={2022},
  publisher={ACM New York, NY}
}

@article{mathews2006nudged,
  title={Nudged elastic band calculation of minimal energy paths for the conformational change of a GG non-canonical pair},
  author={Mathews, David H and Case, David A},
  journal={Journal of molecular biology},
  volume={357},
  number={5},
  pages={1683--1693},
  year={2006},
  publisher={Elsevier}
}

@book{williams2006gaussian,
  title={Gaussian Processes for Machine Learning},
  author={Williams, Christopher K. I. and Rasmussen, Carl Edward},
  year={2006},
  publisher={MIT Press},
  address={Cambridge, MA}
}

@article{riebesell2025framework,
  title={A framework to evaluate machine learning crystal stability predictions},
  author={Riebesell, Janosh and Goodall, Rhys EA and Benner, Philipp and Chiang, Yuan and Deng, Bowen and Ceder, Gerbrand and Asta, Mark and Lee, Alpha A and Jain, Anubhav and Persson, Kristin A},
  journal={Nature Machine Intelligence},
  volume={7},
  number={6},
  pages={836--847},
  year={2025},
  publisher={Nature Publishing Group UK London}
}

@misc{facebookresearch_fairchem,
  author = {{FAIR Chemistry} team},
  title = {{FAIRChem}: FAIR Chemistry — library of machine learning methods for chemistry},
  howpublished = {\url{https://github.com/facebookresearch/fairchem}},
  note = {commit hash as of accessed version, accessed 2025-12-09},
  year = {2025}
}

\newpage

\section*{Supplementary Information}

\subsection*{Modeled Forces}

\begin{figure}[h!]
\centering
\includegraphics[width=1.0\textwidth]{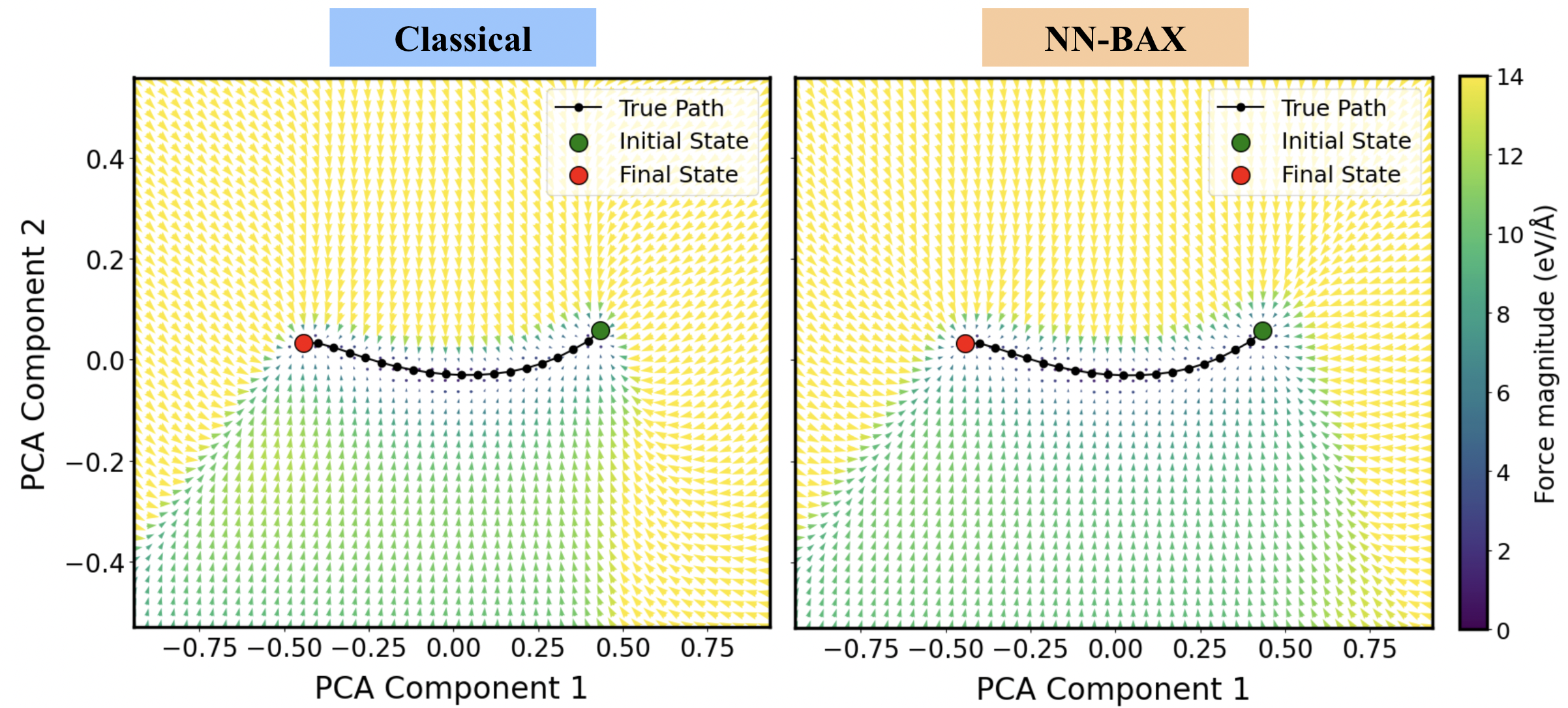}
\caption[Forces]{Modeled Forces for $LJ_7$ system with 0 minima. The forces are 21-dimensional, so the first two principal components are displayed. The left "classical" plot shows the true forces. The right plot shows the modeled forces from NN-BAX, at the final converged iteration.}
\label{fig:7}
\end{figure}

\subsection*{Computational Discussion}
In order to calculate the time per DFT calculation NN-BAX would require to be faster than classical NEB, we run another test on the $LJ_{38}$ 0 minima path to estimate the computational overhead of BAX. NN-BAX took 9.3 hours to run, for 30 BAX iterations, with 200 NEB steps, and with each model trained for 50 epochs. All of our code was run on a single NVIDIA Tesla A100 GPU. Since calls to the LJ potential are practically instant, this is a good approximation for solely the computational overhead cost of NN-BAX. Thus for a setting in which DFT is being used, for NN-BAX to achieve a wall-clock speedup relative to Classical NEB, each DFT simulation would have to be longer than 25.4 seconds. However our setup involves minimal implementation overhead, and can greatly be improved. 

The two primary components of the overhead are model inference during NEB, and training the model on sampled points. In our implementation these components took similar times, with training taking 48\% of the time. Due to the default ASE NEB implementation, images are evaluated sequentially in our NEB runs. Taking advantage of GPU parallelism for model inference would cut the NEB run time down by a factor of 20. For the training component, we retrain the model from scratch at each iteration, with a training budget of 50 epochs. Strategies to reduce this include using a warm start, where we may learn in some fraction of those iterations, like 10 epochs. Taking into account a potential factor of 20 speedup for model inference during NEB and a factor of 5 speedup for training results in a time of 3.1 seconds per DFT simulation, in order to achieve a speedup relative to classical NEB. We note that we can achieve an even greater speedup by using a smaller EquiformerV2 model (see below).

\subsection*{Transition State Predictions}
\begin{figure}[ht]
\centering
\includegraphics[width=1.0\textwidth]{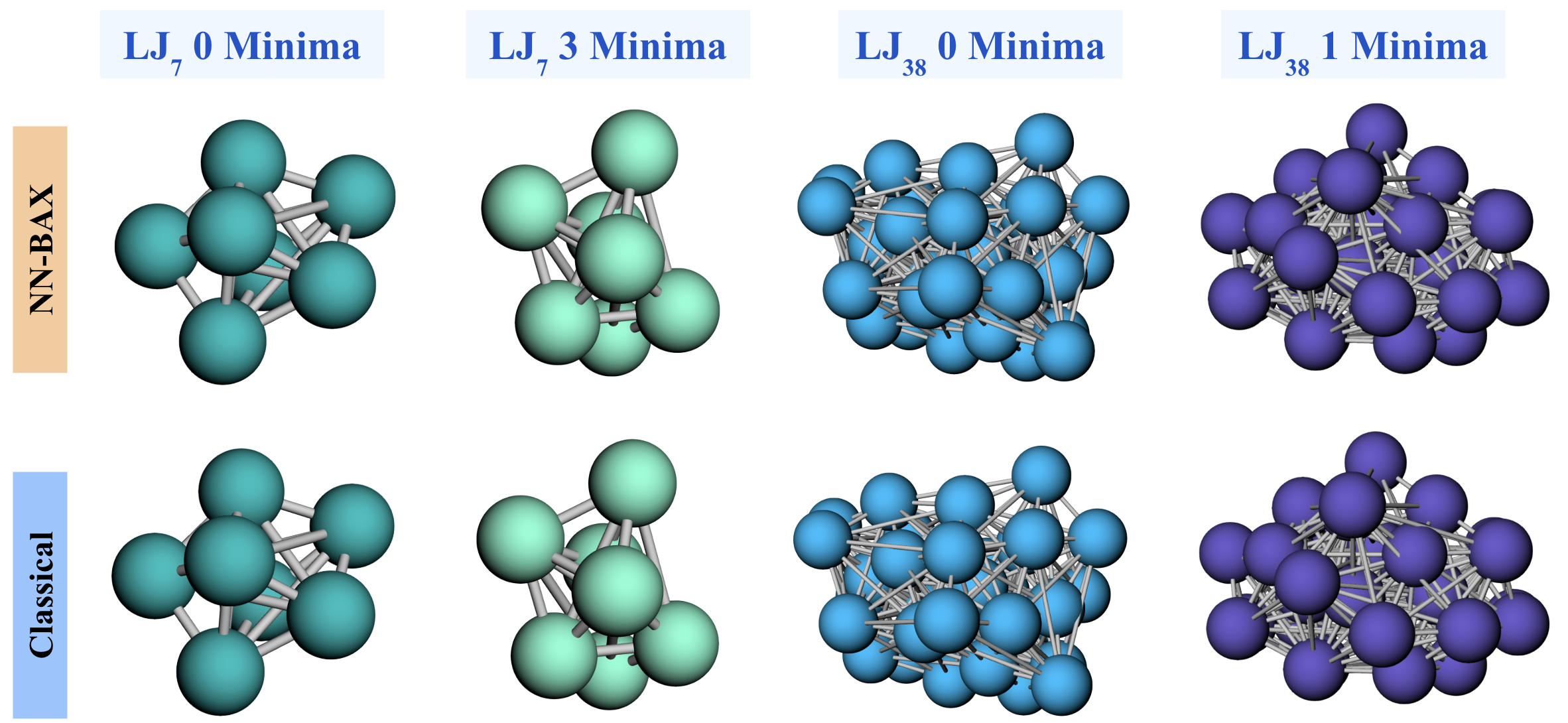}
\caption[Transitions]{Transition state predictions. The NN-BAX predictions and the classical ground truth atomic arrangements are displayed. }
\label{fig:8}
\end{figure}

Minima of potential energy surfaces have a zero gradient in all dimensions. Transition states are a minimum in all dimensions except for one, and correspond to a maximum of potential energy along a reaction path. In three dimensions for example, a saddle point satisfies this constraint. A transition state is often of particular interest because it dictates the activation energy of the reaction and corresponds to bonds breaking and new bonds forming~\cite{eyring1935activated}. Transition states are helpful in understanding the kinetic properties of chemical systems, such as reaction rate. In Figure~\ref{fig:8} we display the predicted structures for the highest energy transition state of each reaction. We observe the NN-BAX structures qualitatively match the classical NEB structures. 

\subsection*{NN-BAX Convergence}

\begin{figure}[t]
\centering
\includegraphics[width=0.85\textwidth]{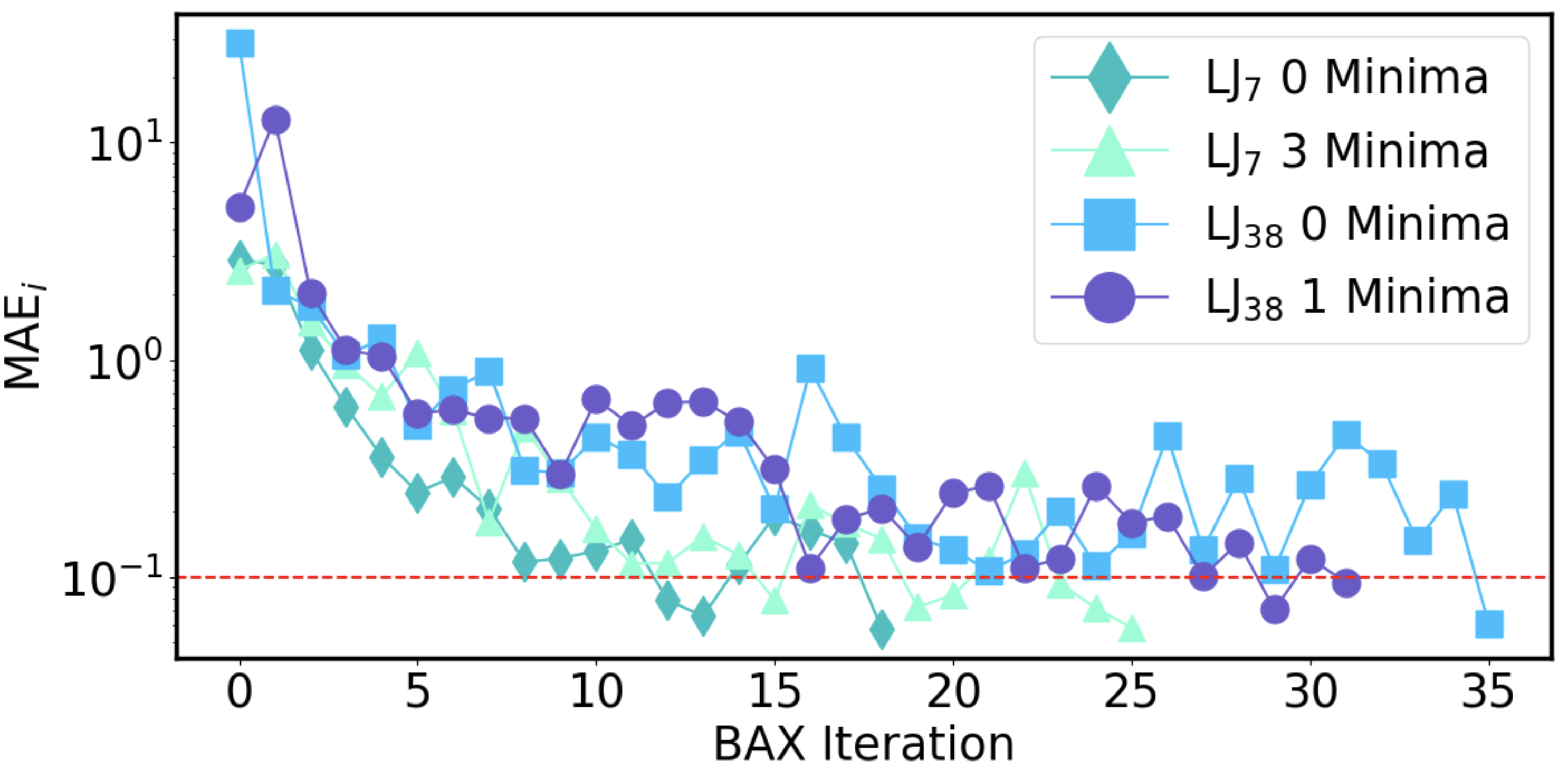}
\caption[Convergence]{$\mathrm{MAE}_i$ convergence metric versus BAX iteration. The red line denotes the convergence threshold of 0.1}
\label{fig:9}
\end{figure}

\begin{figure}
\centering
\includegraphics[width=1\textwidth]{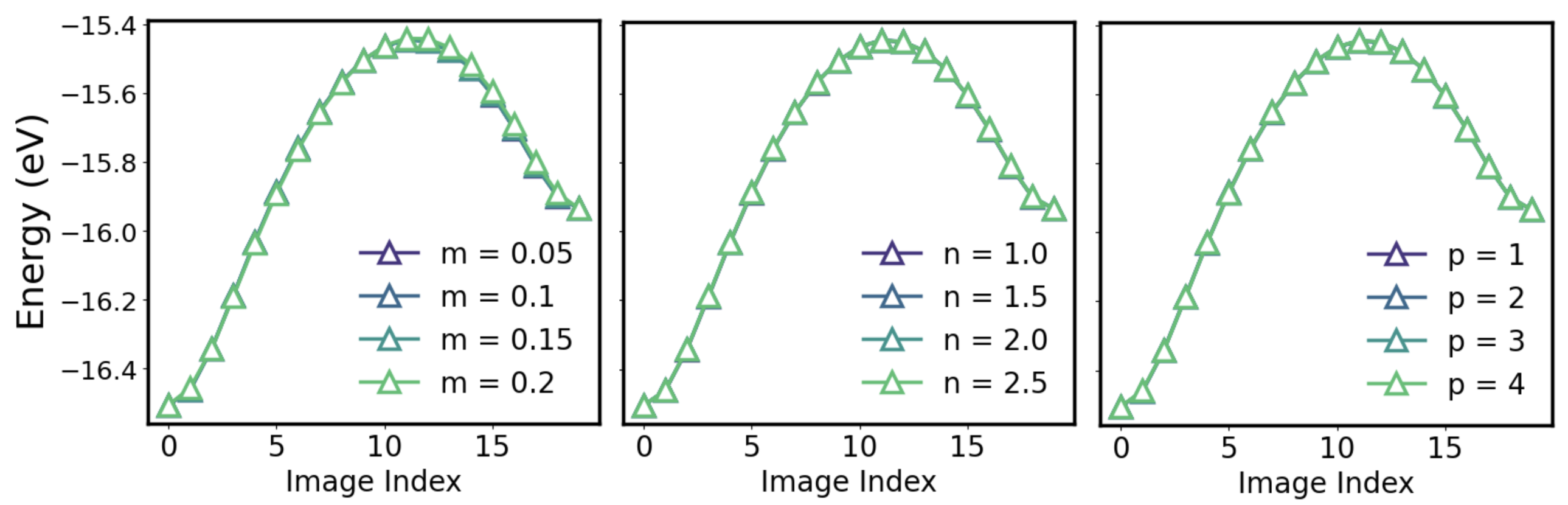}
\caption[Robust]{Probing robustness of NN-BAX convergence hyperparameters for the $LJ_7$ system with 0 minima. For each plot we vary the one of the default values of $m=0.1$, $n=2$, and $p=2$ and plot the resulting energy profile. }
\label{fig:10}
\end{figure}

In Figure~\ref{fig:9} the $\mathrm{MAE}_i$ convergence parameter for the four LJ paths we study is plotted. We observe that it is stable, and reliably falls under the threshold of 0.1 when NN-BAX is outputting the correct path. We note that in the Foundation-BAX study, lower $\mathrm{MAE}_i$ thresholds are required. Specifically, we use $\mathrm{MAE}_i$ = [0.1, 0.1, 0.025, 0.05, 0.025] for paths from points A through F respectively, observing that lower convergence thresholds suit the later paths better. We also observe slightly higher patience values of [4, 2, 2, 3, 3] which suit Foundation-BAX better. 
For the EAM study $\mathrm{MAE}_i$ = 0.001 was found to be optimal, likely because the EquiformerV2 model was already more accurate before any fine-tuning. In Figure~\ref{fig:10} we comprehensively test the robustness of each of the convergence hyperparameters $m$, $n$, and $p$, observing that NN-BAX returns the correct energy profile for various values of these hyperparameters.

\subsection*{Ablation Study}
\begin{figure}[t]
\centering
\includegraphics[width=1.0\textwidth]{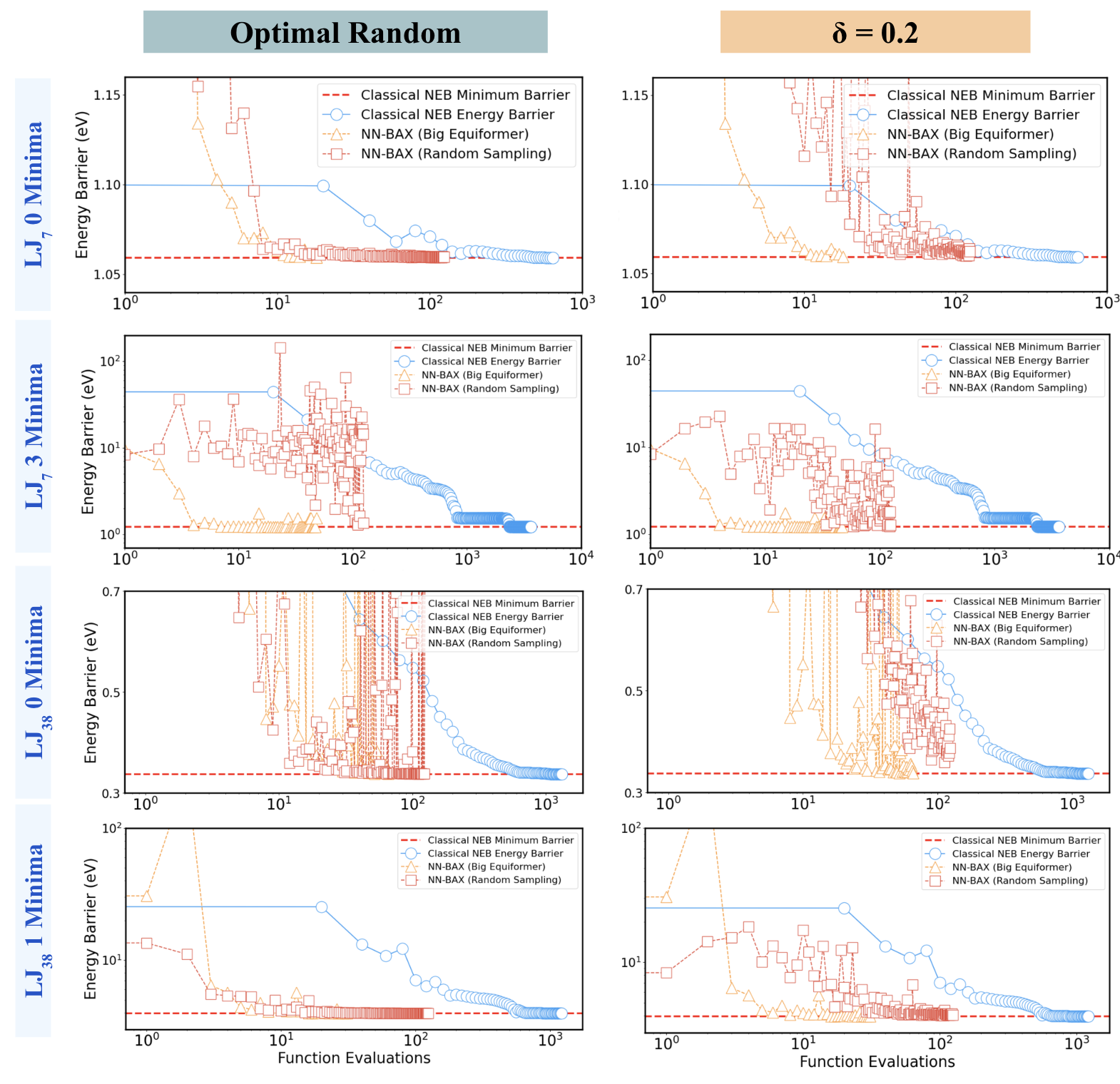}
\caption[rand]{Ablation study with acquisition method, comparing NN-BAX against random sampling method, keeping the fine-tuning procedure the same. We analyze runs with an optimal sampling variation $\delta$ and with $\delta$ = 0.2. }
\label{fig:11}
\end{figure}

We perform an ablation study in order to assess the strength of our acquisition strategy (sampling randomly from the algorithm output). We contrast our method with a random sampling method, which samples points around the linear interpolation between the initial and final state. $\delta$. Specifically, the random sampling method involves first randomly picking a image along the linear interpolation, then varying all atomic coordinates according to a uniform distribution $\in[-\delta, +\delta]$. We compare two methods of random sampling, one where we simply set $\delta$ = 0.2, and one where we use the true path to compute some optimal $\delta$, given by the distance the transition state is from the linear interpolation. We run the random sampling method for 125 BAX iterations and plot the results. We observe that for $\delta$ = 0.2, the random sampling method consistently performs worse, roughly an order of magnitude. For the optimal sampling method we observe better performance, with the performance being on par with NN-BAX for some paths, but still being slower and unable to find the true path for others. However this optimal sampling method is unrealistic as it assumes knowledge of the true path transition state, which is not known prior to running NEB. The results are shown in Figure~\ref{fig:11}.

\subsection*{Smaller EquiformerV2}
\begin{figure}[t]
\centering
\includegraphics[width=1.0\textwidth]{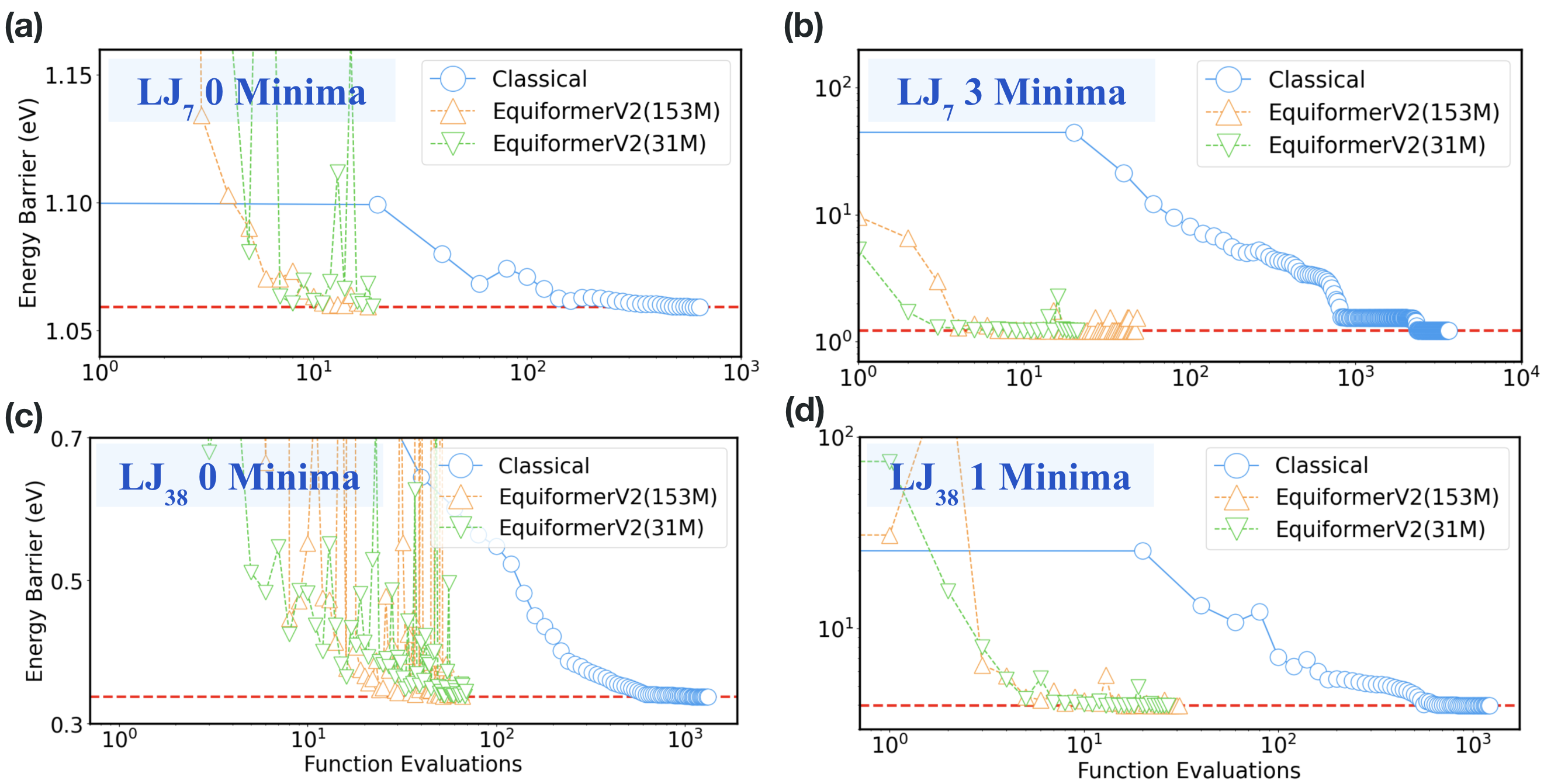}
\caption[Small]{Smaller EquiformerV2 comparison. The EquiformerV2 model with 31M parameters is compared against the one with 153M parameters.}
\label{fig:12}
\end{figure}

In Figure~\ref{fig:12} we compare a smaller EquiformerV2 model (31M params) against the model we used (153M params). We observe generally similar performance. Furthermore, the smaller EquiformerV2 has a faster inference and training time. Specifically, we observe a 1.6$\times$ speedup factor for training, and a 3.8$\times$ speedup factor for inference. 

\end{document}